\documentclass[fleqn,usenatbib]{mnras}

\usepackage[T1]{fontenc}

\DeclareRobustCommand{\VAN}[3]{#2}
\let\VANthebibliography\thebibliography
\def\thebibliography{\DeclareRobustCommand{\VAN}[3]{##3}\VANthebibliography}

\usepackage{graphicx}	
\usepackage{amsmath}	
\usepackage{amssymb}	
\usepackage{mathtools, cuted}
\usepackage{bm}

\defcitealias{soyuer2021}{S21}
\defcitealias{zwick2022}{Z22}
\usepackage{newtxtext,newtxmath}

\title[Localising Planet 9 with the Uranus Mission]{Prospects for localising Planet 9 with a future Uranus mission}

\author[Bucko et al.]{
Jozef Bucko,\thanks{E-mail: jozef.bucko@uzh.ch}
Deniz Soyuer
and Lorenz Zwick
\\
Center for Theoretical Astrophysics and Cosmology, Institute for Computational Science, University of Zurich, Winterthurerstrasse 190, CH-8057 Zurich
}

\date{Accepted 2023 June 02. Received 2023 June 02; in original form 2023 March 29}

\pubyear{2023}

\begin{document}
\label{firstpage}
\pagerange{\pageref{firstpage}--\pageref{lastpage}}
\maketitle
\begin{abstract}
Past years have seen various publications attempting to explain the apparent clustering features of trans-Neptunian objects, the most popular explanation being an unconfirmed "Planet 9". The recently proposed Uranus Orbiter and Probe mission by NASA's \textit{Planetary Science and Astrobiology Decadal Survey} could offer the opportunity to precisely determine Planet 9's sky location and mass by carefully monitoring ranging data during the interplanetary cruise. We use Monte Carlo-Markov Chain methods to reconstruct simulated spacecraft trajectories in a simplified solar system model containing Planet 9, providing an estimate of the  mission's localisation capacity depending on sky location, Earth-spacecraft Doppler link noise level and data collection rate.
We characterise the noise via the Allan deviation $\sigma_{\rm A}$, scaled to the \textit{Cassini}-era value $\sigma_{\rm A}^{\rm \scriptscriptstyle Cass} = 3 \times 10^{-15}$, finding that daily measurements of the spacecraft position can lead to $\sim$0.2 deg$^2$ localisation of Planet 9 (assuming $M_9 = 6.3 M_{\oplus}$, $d_9 = 460$AU). As little as a 3-fold improvement in $\sigma_{\rm A}$ drastically decreases the sky localisation area size to $\sim$0.01 deg$^2$.
Thus, we showcase that a future Uranus mission carries a significant potential also for non-Uranian science.
\end{abstract}




\section{Introduction}

\begin{figure}
    \centering
    \includegraphics[width = \columnwidth]{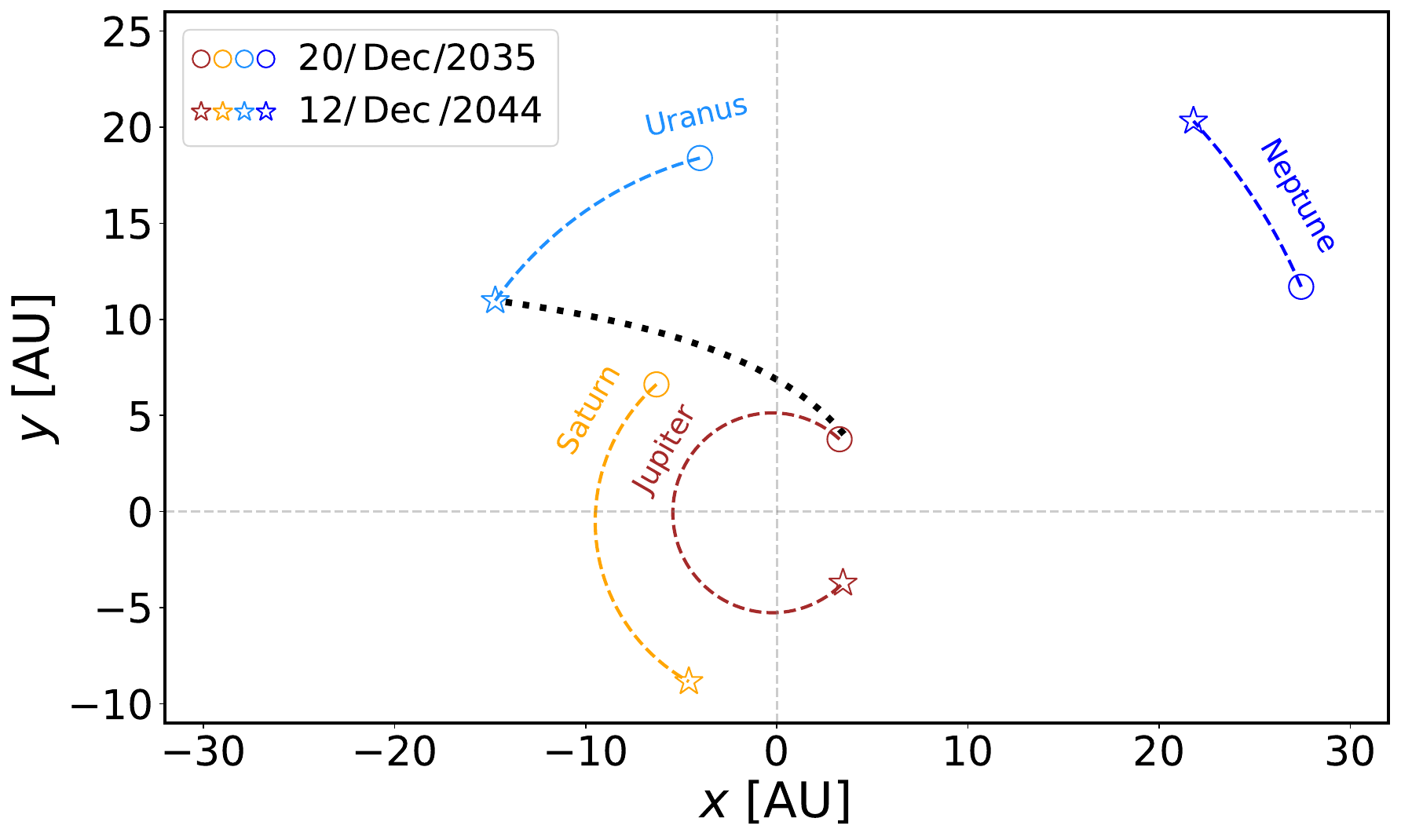}
    \caption{A plausible trajectory of the Uranus Orbiter and Probe mission 
    in the International Celestial Reference Frame. The mission timestamps are provided in the publicly available \href{https://drive.google.com/file/d/1TxDt_qU6H2j2fYGqcDUTJQioSJ2W_KnN/view?usp=sharing}{mission document}. The dashed line denote the orbits of the solar system giants, while the circles and the stars correspond to their location at the given timestamps.}
    \label{fig:mission}
\end{figure}

Since its initial proposal in \citet{planet9}, several papers have addressed possible detection methods for Planet 9  or Telisto (hereafter P9) \citep[see e.g.][]{fienga, p9_cassini, iorio, p9_atacama, planet9_pop, p9_iras}, its origin \citep{p9_captured, p9_50}, and its gravitational influence on other solar system bodies \citep{p9_inclination, p9_hyperbolic_comets, p9_2022}. In this letter, we investigate the potential of the recently proposed Uranus Orbiter and Probe mission\footnote{Survey provided in: \href{https://science.nasa.gov/science-pink/s3fs-public/atoms/files/Decadal-Strategy-Planetary-Science-and-Astrobiology-2023\%E2\%80\%932032.pdf}{Origins, Worlds, and Life: A Decadal Strategy for Planetary Science and Astrobiology 2023-2032 (2022)}}  to constrain the sky position, distance and mass of a hypothetical P9 by examining the ranging data of its radio link with Earth. A plausible mission trajectory is depicted in Figure \ref{fig:mission}, in which the spacecraft undergoes a Jupiter Gravity Assist before entering a 9 year cruise phase to the Uranian system. In particular, our goals are to provide a proof-of-concept of the trajectory reconstruction method detailed in Section \ref{sec:methods} and to estimate the mission's sky localisation accuracy as a function of the ranging noise and data collection frequency in Section \ref{sec:results}. \newline

This letter is a companion piece to the previous works highlighting the scientific potential of collecting ranging data throughout the long cruise time of a prospective Uranus mission (hereafter denoted as PUM), which includes the possible detection of gravitational waves in the mHz frequency range \citep[][hereafter \citetalias{soyuer2021}]{soyuer2021} as well as  constraints on the dark matter content in the solar system \citep[][hereafter \citetalias{zwick2022}]{zwick2022}.

\section{Methods} 
\label{sec:methods}
\subsection{Doppler tracking}
The trajectories of interplanetary spacecraft are tracked via the Doppler time series of the Earth--spacecraft radio link (see also \citetalias{soyuer2021} and \citetalias{zwick2022} for a more detailed overview in this context).
Ranging data is often expressed in terms of the two-way frequency fluctuation $y_2 = \Delta f/f_0$, where the Doppler shift $\Delta f$ is normalised by the link's carrier frequency $f_0$. Thus, the ranging uncertainty on the spacecraft's radial velocity $v_{\rm r}$ can be expressed as $\Delta v_{\rm r} = c \Delta y_{2}$, where $\Delta y_2$ is the noise on the frequency fluctuation, and $c$ the speed of light. 
In the steps of \citetalias{soyuer2021} and \citetalias{zwick2022}, we use the nominal values of the \textit{Cassini} mission as our baseline, and express the noise of the frequency fluctuation as a scaling law with respect to $f_0$, Allan deviation $\sigma_{\rm{A}}$ and two-way light travel time $T_2$ \citep[see also][]{comoretto}:
\begin{equation}
\label{eq:freqfluc}
    \Delta y_2 \approx 6\times 10^{-13} \times \frac{f_0^{\rm \scriptscriptstyle Cass}}{f_0} \frac{\sigma_{\rm{A}}}{\sigma_{\rm A}^{\rm \scriptscriptstyle Cass}}\sqrt{\frac{T_2}{T_2^{\rm \scriptscriptstyle Cass}}}.
\end{equation}
Here $\sigma_{\rm A}^{\rm \scriptscriptstyle Cass} \approx 3 \times 10^{-15}$ and $T_2^{\rm \scriptscriptstyle Cass} = 5730$s. Equation (\ref{eq:freqfluc}) suggests that the noise on the  frequency fluctuation  can be improved either by reducing the Allan deviation or upgrading the  link to higher frequencies (see \citet{armstrong} for an extensive review of noise improvements, and \citet{{optick1,optic0}} for prospects of using optical instruments for reducing atmospheric phase noise). For the remainder of this work, we assume that the link operates at the Ka-band ($\sim$32 GHz) and express our results solely as a function of Allan deviation. For a detailed explanation regarding the frequency fluctuation, we refer the reader to \citetalias{soyuer2021} and \citetalias{zwick2022}.

\subsection{Order of magnitude estimates}
It is useful to go through simplistic order of magnitude calculations first in order to estimate the feasibility of determining P9's parameters.
The acceleration $A_9$ induced on a spacecraft due to the gravitational influence of an additional body can be estimated as
\begin{equation}
    A_9 \approx GM_9 / d_9^2 = 5.3 \times 10^{-13}~~ {\rm m/s^2},
\end{equation}
with the expected values $M_9 \sim 6.3 M_{\oplus}$ and $d_9 \sim 460$AU of P9 \citep{planet9_pop}.
Thus, a spacecraft in the solar system would be displaced by roughly $\sim$20 km from its projected trajectory within a period of $\sim$10 yr. 
Considering that ranging measurements with uncertainty below $0.1$~mm/s throughout the mission have already been accomplished by the New Horizons mission, enabling to track the spacecraft position with a precision of $\mathcal{O}(10{\rm m})$ at 50AU  \citep[see e.g.][with technologies that were proposed a dozen years ago]{newhor08,newhor13}, the prospect of a precise measurement of P9's parameters is realistic.

Recently, \citet{planetx} have analysed the effect of P9's induced tidal gravitational field on other solar system planets via the ranging measurements of \textit{Juno}, \textit{Cassini}, and Mars-orbiting spacecraft. They find that a $5\sigma$--detection of a $5 M_{\oplus}$ Planet at 400 AU would be achievable over the full sky but over only $5\%$ of the sky at 800 AU. Upcoming Mars ranging missions do not improve these estimates significantly, since the degeneracy with the Kuiper Belt's fluctuating quadrupole moment creates a measurement bottleneck. Additionally, the aforementioned work examines the effect of P9's tidal field on Jovian Trojans and its traceability by the Vera C. Rubin Observatory (VRO). Following the order of magnitude  estimation in \citet{planetx}, where Uranus is tracked via the PUM's orbiter, assuming $M_9 = 5 M_\oplus$, $d_9 = 400 \rm{AU}$ and a reasonable 4 year permanence in orbit, the maximum radial displacement in Uranus' orbit is $\delta x \approx 200 \rm{m}$. Thus, the angular shift would correspond to
\begin{equation}
    \Delta \theta = \frac{\Delta x}{d_{\rm{U}}} = \frac{220 ~\rm{m}}{19.8 ~\rm{AU}}\approx 14 \mu \rm{as} 
\end{equation}
which is $\sim$10 times smaller than what it is calculated for Jovian Trojans with 12 years of VRO tracking \citet{planetx}. Thus, in this work we focus on directly reconstructing P9's gravitational influence on the spacecraft trajectory during its interplanetary cruise rather than when it is on the orbit of Uranus.

\subsection{Trajectory reconstruction method}
\label{sec_sub:mcmc_setup}
In order to differentiate the gravitational effects of P9 on the spacecraft from those of other solar system bodies, we developed a numerical procedure based on a Monte Carlo--Markov Chain (MCMC), in which we reconstruct the gravitational influence of P9 in the ranging data of thousands of simulated PUMs (see also \citetalias{zwick2022}). The tracking of each virtual spacecraft begins from the moment it exits Jupiter's sphere of influence and ends when it reaches Uranus. Its trajectory is solely influenced by graviational forces and is calculated with a symplectic integrator, ensuring the conservation of energy. The spacecraft cruise within a simplified solar system model consisting of the Sun, Jupiter, Saturn, Uranus and Neptune and P9. The positions of the outer planets are not integrated, but rather sampled at the required time-steps from the JPL HORIZONS database using the \texttt{astroquery} tool by \citet{query}. Therefore, the 9 free parameters considered in our simulations are the masses of the Sun and outer planets along with the standard gravitational parameter, distance, right ascension $\alpha$ and declination $\delta$ of P9.

Our virtual ranging data consists of the frequency fluctuation $y_2$ received from each a spacecraft realization along its trajectory. Denoting a specific 9-dimensional configuration of our simplified solar system $\theta$, we accommodate a log-likelihood in the form of
\begin{equation}
     \log \mathcal{L (\theta)} = -\frac{1}{2}\left( \xi(\theta)  - \xi_{ \rm   GT}   \right) \Sigma^{-1} \left( \xi(\theta) - \xi_{\rm GT} \right)^T,
\end{equation}
where $\xi(\theta)$ is a ranging data vector received along the spacecraft cruise for a given parameter space configuration $\theta$ and for an assumed data collection rate of the signal. The subscript GT denotes the ground truth signal obtained  from a trajectory defined by the fiducial solar system configuration $\theta_{\rm fid}$, which is a fixed value throughout a specific MCMC run. Similar to \citetalias{zwick2022}, we consider $\sigma_{\rm A}$ to be descriptive of the dominant noise sources on the Doppler link signal and therefore define the covariance matrix of the observed signal as $\Sigma := {\rm diag} (\sigma_{\rm A}^2)$. Additionally, we perform our analysis  comparing different data collection rate of received Doppler link signal; probing once-per-day, once-per-week and once-per-month configurations.

To make sure our results are not prior-dominated, we sample $\Delta G M$ from a flat prior that is eight times wider than the recent limits on the  standard deviations of $GM$ for chosen objects. The fiducial values as well as the standard deviations ($\sigma_0 = 9 \times 10^9 \rm m^3 s^{-2}$) are taken from \citet{park21}. Additionally, we sample over P9's standard gravitational parameter $GM_9 \in [GM_\oplus, 10 GM_\oplus]$, its radial distance from the Sun $d_9\in [200{\rm AU},1000{\rm AU}]$ and the right ascension $\alpha$ and declination $\delta$ of P9 on the sky. 
To enhance the MCMC convergence, we initialize our MCMC walkers in the vicinity of the ground truth parameter values using a Gaussian centered on the fiducial value. For the standard gravitational parameters of the Sun and outer planets, the spread of the Gaussian is one standard deviation of $GM$. For P9 parameters, we take $1 GM_\oplus$ for $GM_9$, 50AU for $d_9$ and $\sim$$20^\circ$ for $\alpha$ and $\delta$.  We consider our MCMCs converged once $R_c \leq 1.1$ as defined within Gelman-Rubin criterion \citep[see][for details]{GelmanRubin1992}. 

To perform our MCMC analysis, we use \texttt{emcee} ensemble sampler \citep{emcee} and choose `stretch move' \citep{camcos} as a sampling (move proposal) algorithm. We pa\-rallelise our pipeline using the \texttt{multiprocess} package \citep{multi} by running each MCMC walker on a separate OpenMP thread. Our work makes use of the computational resources of the Swiss National Supercomputing Centre.\footnote{\url{https://www.cscs.ch}}

\begin{figure*}
     \centering  
     \includegraphics[width=\textwidth]{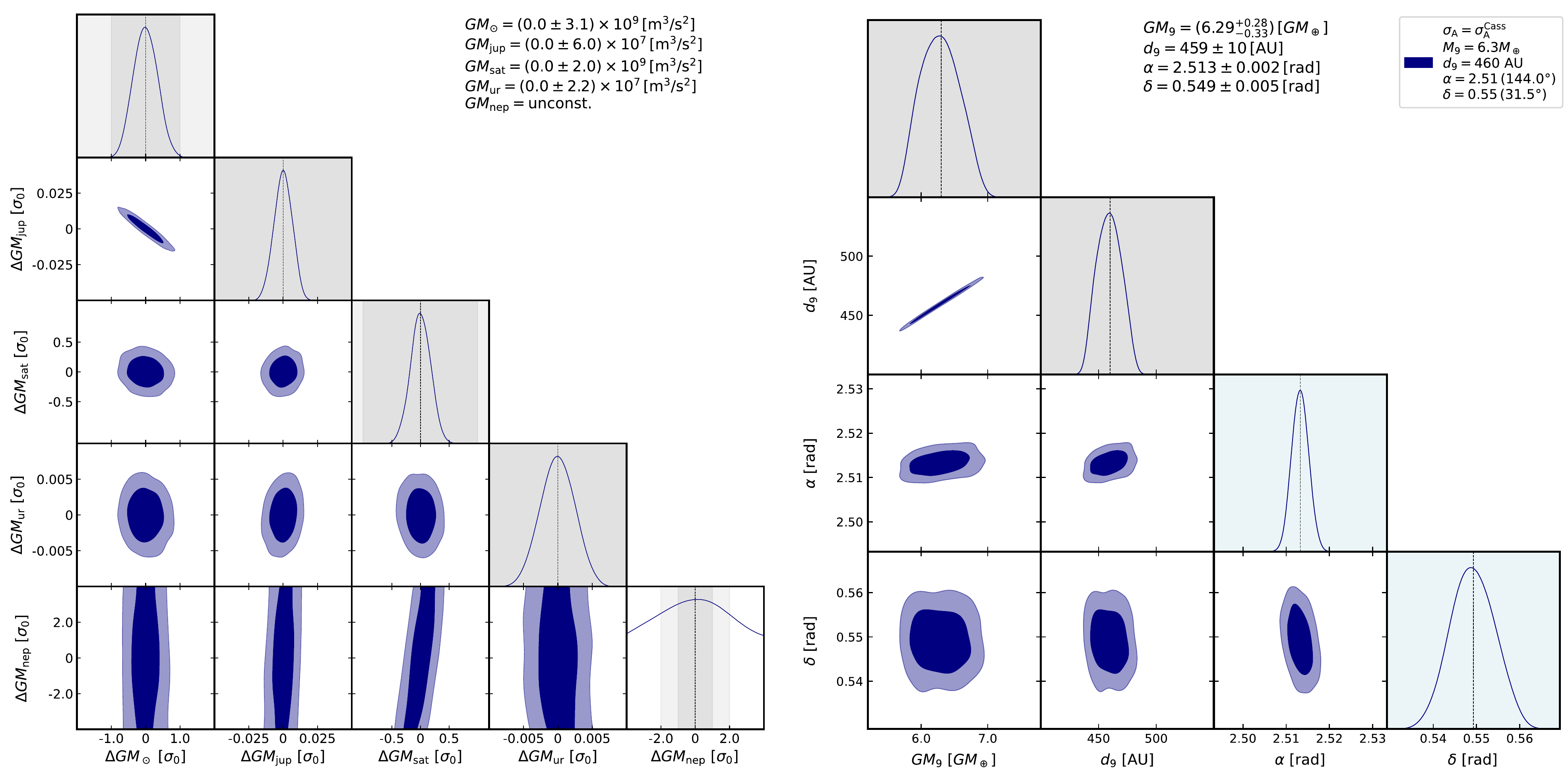}
     \label{fig:mcmc1}
    \caption{Posteriors of the MCMC parameters of the Sun and the giant planets (left panel) and P9 (right panel) for daily ranging measurements of the spacecraft trajectory with $\sigma_{\rm A} = 3 \times 10^{-15}$. The posteriors for the $GM_i$ are displayed in factors of $\sigma_0 = 9\times  10^9 ~ {\rm m^3 s^{-2}}$. For each parameter, the purple contours represent the $1\sigma$ and $2\sigma$ confidence intervals. In the diagonal panels, the dashed lines indicate the positions of the underlying ground truth for the assumed MCMC setup, while shaded bands mark the 1$\sigma$ and 2$\sigma$ observational uncertainties as reported by \citet{park21} for the Sun and outer planets and by \citet{p9_bh} for P9. The light blue regions in the last two diagonal panels display the prior assumed in our MCMC analysis for ($\alpha$,$\delta$).}
        \label{fig:mcmc}
\end{figure*}

\section{Results} 
\label{sec:results}
\subsection{Parameter posteriors}

In Figure~\ref{fig:mcmc} we show an example of the marginalised posteriors as obtained from a specific MCMC run, given a set of fiducial values for the standard gravitational paramaters of the Sun and outer planets.
The position of P9 on the sky is similarly set to ($\alpha$, $\delta$) = (144.0$^\circ$, 31.5$^\circ$). The noise level is normalised to {\it Cassini}-era values and the data collection rate  is set to 1 data point-per-day. With these choices, it is clearly visible that the ground truth values of all varied parameters were recovered reliably and the results discussed below are representative of the large majority of our MCMC runs.

Due to initial and final location of the spacecraft, the method is most sensitive to the standard gravitational parameters of Jupiter and Uranus. Constraints on the latter are improved by a factor of more than 100 compared to recently reported values \citep{park21}, though this estimate is bound to degrade if the Jovian and Uranian system were modelled with more complexity (i.e.accounting for their many satellites). In case of the Sun and Saturn, the obtained constraints are a few times stronger than the ones reported by \citet{park21}, while the ranging data are not very sensitive to the mass of Neptune within the assumed prior. Furthermore, we can observe an expected anti-correlation in masses of the Sun and Jupiter as these two objects mostly contribute to the total mass decelerating the spacecraft along its cruise. The grey shaded regions in the diagonal panels in Figure~\ref{fig:mcmc} represent 1$\sigma$  and 2$\sigma$ bounds on the sampled parameters, as published recently in \citet{p9_bh,park21}, while the light blue regions indicate the 20$^\circ$ domain around the ground truth values of $\alpha$ and $\delta$. Crucially, the standard gravitational parameter of P9 and its distance from the centre of the solar system $d_9$ are found to be constrained by a factor of 10 better that the up-to-date limits obtained by \citet{p9_bh}. There is an expected strong correlation of $GM_9$ and $d_9$, caused by the fact that ranging signal is sensitive to the product $GM_9/d_9^2$ and less so to $GM_9$ or $d_9$ separately. 

\subsection{Sky localisation}
\begin{figure*}
    \centering
    \includegraphics[width = 0.92\textwidth]{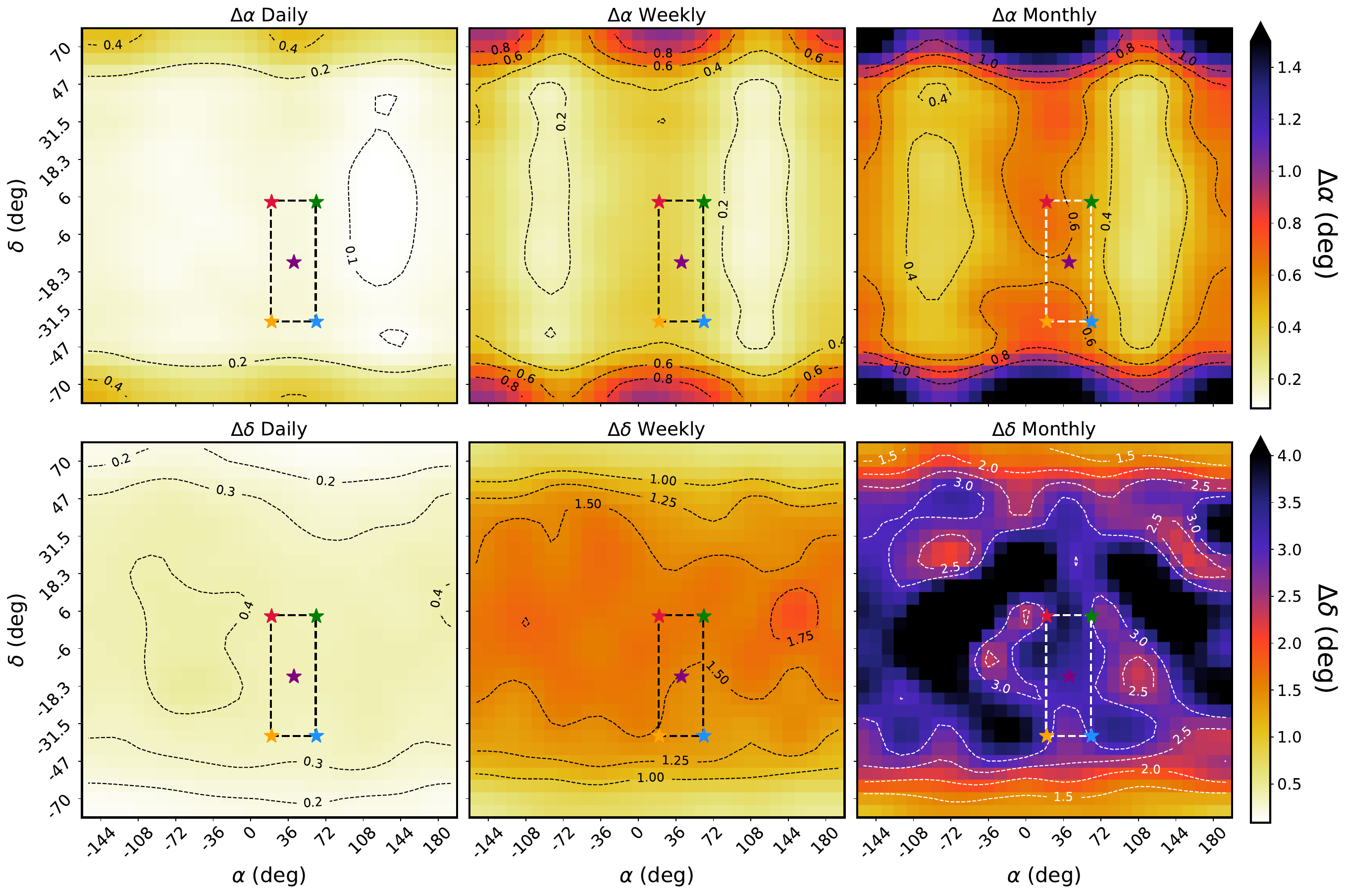}
    \caption{Uncertainties in the right ascension $\alpha$ (top panels) and declination $\delta$ (bottom panels) of P9's position. Left, center and right panels show the recovered assuming a \textit{Cassini}-era Allan deviation of $\sigma_{\rm A} = 3 \times 10 ^{-15}$ and a daily, weekly or monthly data collection rate, respectively. The coloured stars enclose the area on the sky that is likely to harbor P9 according to \citet{p9_cassini}, i.e. $\pm 20^\circ$ around $(\alpha, \delta) = (40^{\circ}, -15^{\circ})$.}
    \label{fig:uncert}
\end{figure*}
We perform a sensitivity study with respect to angular position of P9 on the sky by running run a two-dimensional grid of MCMCs with varying right ascension $\alpha \in [0,2 \pi]$ and declination $\delta\in[-\pi/2,\pi/2]$ angles.
For every $(\alpha,\delta)$ configuration we infer the width of the recovered posteriors $(\Delta \alpha,\Delta \delta)$, which we interpret as the final uncertainty of P9's localisation. In the left column of Figure~\ref{fig:uncert}, we visualize the two-dimensional parameter space of right ascension (top panel) and declination (bottom panel). Each pixel in the panels represents a single MCMC analysis with color scale highlighting the recovered angular uncertainties, which are marginalised over the remaining parameters. We find that the baseline scenario assuming once-per-day ranging measurements and {\it Cassini}-era noise levels ($\sigma_A^{\rm \scriptscriptstyle Cass} = 3\times 10^{-15}$) leads to a localisation precision of $\Delta \alpha  \lesssim 0.1 \,\rm deg$ and $\Delta \delta  \lesssim 0.3 \,\rm deg$ in the vast majority of parameter space. The constraints for the right ascension degrade strongly for polar configurations, as expected from the geometry of the system. 

We investigate the effect of data sparsity by repeating the analysis for a data collection rate  of one week and one month, displayed in the central and right columns of Figure~\ref{fig:uncert}, respectively. As expected, the localisation degrades with a reduced number of observational data points. We find the following results: $\Delta \alpha  \lesssim 0.3 \,\rm deg$, $\Delta \delta  \lesssim 1.5 \,\rm deg$ in case of weekly  and  $\Delta \alpha  \lesssim 0.5 \,\rm deg$, $\Delta \delta  \lesssim 3.0 \,\rm deg$ in case of monthly ranging measurements in the large majority of ($\alpha$,$\delta$) parameter space. We observe that the uncertainty in the declination is more sensitive to the data collection rate  of the measured data in comparison to the right ascension. We further determine the effects of varying noise levels. We start with the baseline Allan deviation value of $\sigma_A^{\rm \scriptscriptstyle Cass}$ and subsequently assume improvements in $\sigma_{\rm A}$ by a factor of 3, 10 and~30. We probe the dependency of the localisation constraints on the Allan deviation at five different locations on the sky; a point located at ($40^\circ$,$-15^\circ$) and four additional points deviating by $\pm 20^{\circ}$ (denoted by the coloured stars in Figure \ref{fig:uncert}). These choices are suggested by the works of \citet{fienga, p9_cassini}, in which \textit{Cassini} ranging data is used to constrain P9's position by monitoring perturbations in Jupiter's orbit. In Figure~\ref{fig:uncert_vs_sigma}, we plot the width of the inferred MCMC posteriors (at $1\sigma$ level) for both $\alpha$ (left panel) and $\delta$ (right panel) as a function of the underlying Allan deviation and data collection rate. The individual results at the specified sky locations are averaged and fitted by linear regression, yielding the following power-law relations:
\begin{equation}
    \Delta  \sim k \times \left(\frac{\sigma_{\rm{A}}}{\sigma_{\rm A}^{\rm  \scriptscriptstyle Cass}}\right)^{b_1} \times \left(\frac{T_{\rm{res}}}{1\, \rm{day}} \right)^{b_2},
\end{equation}
where $T_{\rm{res}}$ is the data collection rate in the ranging data, and $(k = 0.15^{{\circ}} \pm 0.01^{{\circ}}, b_1=1.04 \pm 0.08, b_2 = 0.40\pm 0.01)$ for $\Delta \alpha$ and $(k = 0.36^{{\circ}}\pm 0.03^{{\circ}}, b_1=1.51 \pm 0.07, b_2 = 0.71\pm 0.04)$ for $\Delta \delta$. Thus, our baseline result is that a PUM has the potential to constrain the sky localisation box of P9 with an area of approximately 0.2 square degrees at 1$\sigma$, provided P9 is located within the prior found in \citet{fienga, p9_cassini}. The localisation precision improves rapidly with a reduced Allan deviation, scaling as $\sim \sigma_{\rm{A}}^{2.5}$, and degrades approximately linearly with the data collection rate of ranging data, scaling as $\sim T_{\rm{res}}^{1.1}$ .

\section{Discussion and Conclusion}
\label{sec:disc}
\begin{figure*}
    \centering
    \includegraphics[width = 0.91\textwidth]{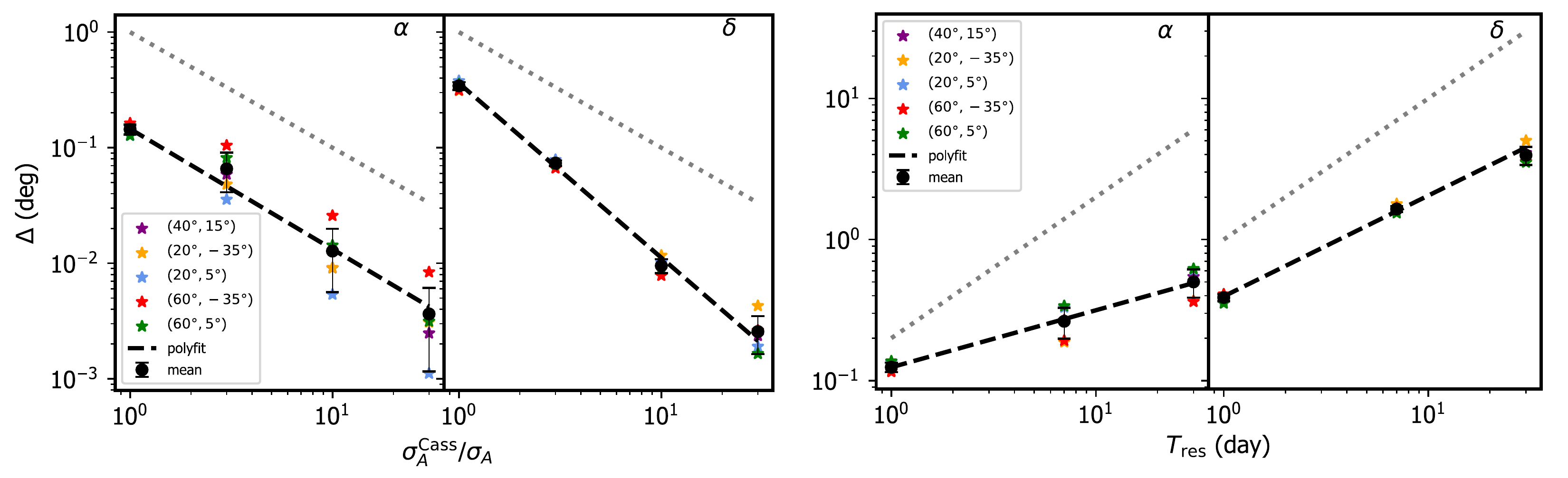}
    \caption{Uncertainty in right ascension $\alpha$  and declination $\delta$  as a function of the Allan deviation (left panels, normalised with the \textit{Cassini}-era value of $3\times10^{-15}$) and the data collection rate  (right panels, normalised with the baseline value of 1 day). The results are evaluated for a set of sky locations in the prior of \citet{p9_cassini} (coloured stars, see also Figure \ref{fig:uncert}). The grey dotted lines guide the eye as to a linear scaling. The black dashed lines denote the best-fit power law to the averaged uncertainties at a given Allan deviation or data collection rate.}
    \label{fig:uncert_vs_sigma}
\end{figure*}
In this work, we showed that by examining ranging measurements during its lengthy interplanetary cruise, the proposed Uranus Orbiter and Probe mission has the potential to  pinpoint the location of the hypothetical P9 on the sky. We performed an extensive sensitivity study, finding that the localisation area of P9 can be constrained to:
\begin{align}
    \sim 0.2\, \rm{deg}^2 \times \left(\frac{\sigma_{A}}{\sigma_{\rm A}^{\rm \scriptscriptstyle Cass}}\right)^{2.5} \times \left(\frac{T_{\rm{res}}}{1\, \rm{day}}\right)^{1.1}
\end{align}
at 1$\sigma$ confidence level, where the Allan deviation is scaled with the \textit{Cassini}-era value of $3\times 10^{-15}$. Taking this baseline at face value, the mission can improve current state of the art ranging constraints by a very promising factor $\sim$$10^3$ compared to \textit{Cassini} tidal measurements \citep{p9_cassini}, aiding the possibility of a visual follow up with electromagnetic instruments \citep[see e.g.][]{2015brown,2022sedgewick,planet9_pop}.

The reliability of our analysis is limited by two main factors. Firstly, the simplistic model of the solar system, which only includes the gravitational field of the outer planets and the Sun. As mentioned in \citep{kuiper}, smaller objects such as dwarf planets, larger moons or a Kuiper belt with mass at the level of  0.02 $M_\oplus$ should in principle be taken into account to accurately model the effects of P9's gravitational field in the solar system. Nevertheless, the trajectory reconstruction method offers an unique advantage with respect to ranging measurements taken in planetary orbit; the heliocentric radius of the spacecraft changes significantly throughout its cruise in the outer solar system. Most confounding gravitational influences will therefore present a clear time dependence, becoming weaker or stronger as the spacecraft travels from $\sim$5 to 20AU. On the other hand, P9's gravitational field will produce an almost constant acceleration throughout the mission's trajectory, a very clear signature unique to distant objects.
Secondly, our treatment of the radio link noise is similarly simplistic, based primarily on a frequency-independent Allan deviation. In reality, one should include individual models of the various astrophysical and mechanical noise sources, and explicitly exclude data gaps caused by solar conjunctions \citep[see e.g.][]{armstrong}. We attempted to model such effects by exploring the consequence of ranging data sparsity, finding that the localisation does indeed degrade when only infrequent ranging data is available (see Figure \ref{fig:uncert_vs_sigma}).  Nevertheless, we have found that even mild improvements in the Allan deviation with respect to \textit{Cassini}-era technology will have drastic consequences on the constraining potential of the mission. Considering 30 years of technological developments, such improvements are realistic \citep[see e.g.][]{optic0,2021Genova}, provided they become one of the mission priorities (see \citetalias{zwick2022} for a more thorough discussion).

In addition to the data collected during the cruise time, Planet 9's gravitational effect can  potentially  be studied in orbit around Uranus. Indeed it has been shown \citep{euro} that if orbiting along wide elliptical trajectories, Planet 9 can have a measurable impact on the spacecraft trajectory.

We believe that our work reliably showcases the potential of a PUM to constrain P9's position and mass with unprecedented precision, in addition to the possibility of detecting gravitational waves (\citetalias{soyuer2021}), constraining the dark matter content in the solar system (\citetalias{zwick2022}), and dynamically measuring Uranus' angular momentum via the Lense-Thirring effect affecting the Uranian orbiter \citep{euro}.

The mentioned results can serve as motivation to focus on collecting and analysing the high-quality ranging data from the prospective Uranus Orbiter and Probe mission.

\vspace{-0.7cm}
\section*{Acknowledgements}
Authors thank Daniel C. H. Gomes for his valuable commentary. We are grateful to Prasenjit Saha for his resolute and graceful leadership. D. Soyuer is thankful to Lara Hildebrand Rey for her  compassionate support. We acknowledge the support of the \textit{International Space Science Institute (ISSI)} in Bern, Switzerland.
\vspace{-0.7cm}
\section*{Data Availability}
The JPL HORIZONS System is publicly accessible. 

\bibliographystyle{mnras}
\bibliography{example} 





\bsp	
\label{lastpage}
\end{document}